%% file: draft-phiee.tex
\documentclass[prd,twocolumn,showpacs,amsmath,amssymb]{revtex4-1}
\usepackage{amssymb}
\usepackage{amsfonts}
\usepackage{overpic,graphicx}
\usepackage{keyval,graphicx}
\usepackage{textcomp,wasysym,ulem}
\usepackage[perpage,symbol]{footmisc}%
\usepackage{lineno}
\usepackage[CJKbookmarks=true, colorlinks=true,
linkcolor=blue, urlcolor=blue,citecolor=blue]{hyperref}
\usepackage[english]{babel}

\begin{document}

\title{\boldmath Search for rare decay $J/ \psi \to \phi e^+ e^-$}

\input{authors_aug2017}

\begin{abstract}
Using a data sample of $448.1\times10^6$ $\psi(3686)$ events collected at $\sqrt{s}=$ 3.686 GeV
with the BESIII detector at the BEPCII, we search for
the rare decay $J/\psi \to \phi e^+ e^-$
via $\psi(3686) \to \pi^+\pi^- J/\psi $.
No signal events are observed and the upper limit on the branching fraction
is set to be $\mathcal{B}(J/\psi \to \phi e^+ e^-) < 1.2 \times 10^{-7}$ at the 90\% confidence level, which
is still about one order of magnitude higher than the Standard Model prediction.

\end{abstract}

\pacs{13.20.Gd, 13.40.Ks}
\maketitle
\oddsidemargin -0.2cm
\evensidemargin -0.2cm

\section{\bf INTRODUCTION}

\begin{figure}[htbp]
\begin{center}
\includegraphics[width=6cm]{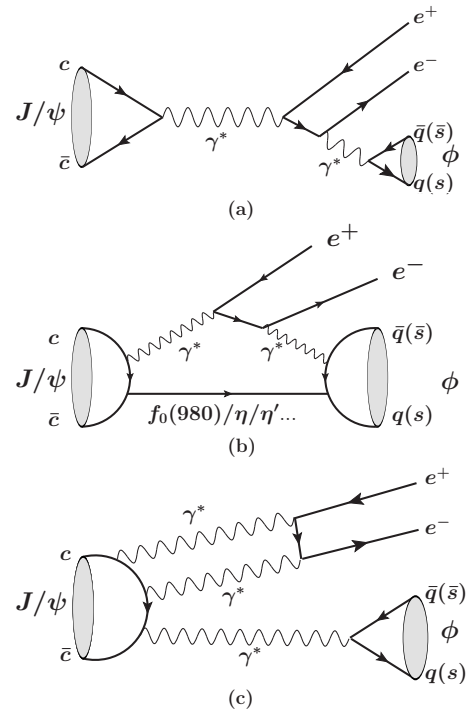}\\
\caption{\small Feynman diagrams contributing to the decay $J/\psi \to \phi e^+e^-$:  (a)~the leading-order EM process, (b)~the EM and strong mixed loop process, and
(c)~the EM process proceeding through three virtual photons. } \label{fig:feynman}
\end{center}
\end{figure}

The BESIII experiment has accumulated $4.48\times10^8$ $\psi(3686)$ events
which is the largest $\psi(3686)$ data sample produced directly in $e^+e^-$ annihilation
in the world currently. By tagging
the two soft pions in the decay of $\psi(3686)\to \pi^+\pi^- J/\psi$,
the final states from $J/\psi$ decay can be well distinguished. This provides an
almost background free sample to investigate the rare $J/\psi$ decay, which may be sensitive to new physics.
The rare decay $J/\psi \to \phi e^+e^-$ is one particulary interesting example~\cite{guoxd}.
This decay channel occurs mainly through the three dynamic processes shown in Figs.~\ref{fig:feynman}(a)-(c).
These include: (a)~the leading-order electromagnetic (EM) process;
(b)~the EM and strong mixed loop process;
and (c)~the EM process proceeding through three virtual photons.
In diagram~(b), the non-perturbative strong loop can be
treated as proceeding through intermediate mesons, as discussed in Ref.~\cite{guoxd}.
Within the framework of the Standard Model~(SM), the partial widths from the leading EM and mixed
loop processes are predicted to be at a level of $10^{-6}$ and $10^{-9}$~keV, respectively,
corresponding to branching fractions at the order of $10^{-8}$ and $10^{-11}$~\cite{guoxd}.
However, if there is a new particle involved in the intermediate process, such as a dark photon
with a mass of several MeV/$c^2$ or a
glueball with certain quantum numbers, contribution from Fig.~1(b) can be enhanced greatly.
Thus, any deviations from the predictions~\cite{guoxd} would hint at the existence of new physics. Alternatively,
if a positive result were obtained with a branching fraction in the expected range, this decay channel
could be used to extract information of some interesting mesons such as $f_0(980)$ or $a_0(980)$ since
their form factors are involved in the predictions.

Although BESIII has also available the currently world's largest data
sample of directly produced $J/\psi$, this is not used in the present
analysis due to badly-controlled background contamination from QED processes.
In this work, we report on search for the rare decay of
$J/\psi \to \phi e^+e^-$ via $\psi(3686) \to \pi^+\pi^- J/\psi$.

\section{BESIII AND BEPCII}
\label{sec:bes3}

The BESIII detector~\cite{bes3} at the BEPCII $e^+e^-$ collider is a major upgrade of
the BESII experiment~\cite{bes2} at the Beijing Electron-Positron Collider~(BEPC)
and is optimized to study physics in the $\tau$-charm energy region.
The design peak luminosity of
BEPCII, $1.0\times10^{33}$~cm$^{-2}$s$^{-1}$ at a center-of-mass energy of $3773$~MeV, was achieved in 2016.
The BESIII detector, with a geometrical acceptance
of $93$\% of the full solid angle, consists of the following five main components.
$(1)$~A small-celled multi-layer drift chamber~(MDC)
with $43$ layers is used for charged track reconstruction and measurement of ionization energy loss ($dE/dx$). The average single-wire
resolution is $135$~$\mu$m, and
the momentum resolution for $1.0$~GeV/$c$ charged particles in a $1$~T magnetic field
is $0.5$\%. The specific $dE/dx$ resolution is 6\% for electrons from Bhabha scattering.
$(2)$~A time-of-flight~(TOF) system surrounds the MDC.
This system is composed of a two-layer barrel, each layer consisting of $88$ pieces
of $5$~cm thick and $2.4$~m long plastic scintillators, as well as two
end caps each with $96$ fan-shaped, $5$~cm thick, plastic scintillators. The time resolution is $80$~ps
in the barrel and $110$~ps in the end caps, providing a $K/\pi$ separation of more than $2\sigma$ for momenta up
to $1.0$~GeV/$c$. $(3)$~An electromagnetic
calorimeter~(EMC) is used to measure photon energies and consists of $6240$ CsI(Tl) crystals
in a cylindrically-shaped barrel and two end caps. For $1.0$~GeV photons, the energy resolution is $2.5$\% in
the barrel and $5$\% in the end caps, and the position resolution is $6$~mm in the barrel and $9$~mm in the end
caps. $(4)$~A superconducting solenoid magnet surrounding
the EMC provides a 1~T magnetic field. $(5)$~The muon chamber system is made of resistive plate chambers with $9$ layers
in the barrel and $8$ layers in the end caps and is incorporated into the return iron yoke of the superconducting
magnet. The global position resolution is about $2$~cm.

Interactions within the BESIII detector are simulated by the {\sc GEANT4}-based~\cite{geant4} simulation software {\sc boost}~\cite{boost},
which includes: geometric and material descriptions of the BESIII detector; detector
response and digitization models;
and a record of detector running conditions and performances.
The production of the $\psi(3686)$
resonance is simulated by the Monte Carlo (MC) generator {\sc KKMC}~\cite{kkmc}, which incorporates the effects of
the energy spread of the beam and initial-state radiation. Known decays are generated by
{\sc EVTGEN}~\cite{evtgen} using the branching fractions quoted by the Particle Data Group (PDG)~\cite{pdg18},
and the remaining unknown decays are generated with the {\sc LUNDCHARM} model~\cite{lundcharm}.

In this analysis, the process $J/\psi \to \phi e^+e^-$ is studied via $\psi(3686) \to \pi^+\pi^- J/\psi$, and
the $\phi$ meson is reconstructed using its decay to $K^+K^-$.
The transition $\psi(3686) \to \pi^+\pi^- J/\psi$ is generated according to
the results of an amplitude analysis as described in Ref.~\cite{bes2jpsipipi}.
The process $J/\psi \to \phi e^+e^-$ is
generated according to the amplitude given in
Ref.~\cite{guoxd}, in which the leading-order EM process is expected to be the dominant
contribution according to the SM prediction. The spin correlation of $J/\psi$ produced in
the previous decay is considered as in Ref.~\cite{lifengyun}.
The decay $\phi \to K^+K^-$ is generated using a $\sin^2\theta_K$
distribution, where $\theta_K$ is the helicity angle of the kaon in the center-of-mass system
of the $\phi$ meson.

\section{Event Selection}
\label{sec:evtslt}

Charged tracks are reconstructed with MDC hits within the range $|\cos\theta|<0.93$, where
$\theta$ is the polar angle with respect to the electron beam direction. They are required to
originate from the interaction region, defined as $R_{xy}<1$~cm and $R_{z}<10$~cm, where $R_{xy}$ and $R_{z}$
are the projections of the distances from the closest approach of the tracks to the interaction point
in the $xy$-plane and in the $z$-direction, respectively.

Particle identification (PID) probabilities for good charged tracks are calculated with the
$dE/dx$ and TOF measurements under the hypothesis that the track originated from a pion, kaon, proton or electron.
For kaon candidates, we require that the probability for
the kaon hypothesis is larger than the corresponding probability for the pion
and proton hypotheses. For electron candidates, the probability for electron hypothesis is
required to be larger than the probabilities for the pion and kaon hypotheses.
To avoid contamination from pions, electron candidates must satisfy the
additional requirement $E/p>0.8$, where $E$ and $p$ represent the energy deposited in the EMC and the
momentum of the electron, respectively.

For the two pions, no PID selection criterion is required. All pairs of opposite charged tracks with
momentum less than $0.45$~GeV/$c$ are assumed to be pions, and their recoil masses $M_{\pi^+\pi^-}^{\rm rec}$ are
calculated and required to be within the range $(3.05, 3.15)$~GeV/$c^2$.

To improve the mass resolution and suppress backgrounds, an energy-momentum
constrained kinematic fit ($4$C) to the initial beam four momentum
is imposed on the selected charged tracks.
The resulting $\chi^2_{4\rm C}$ of the kinematic fit is required to be less
than $40$. If more than one combination is found in an event,
the combination with the least $\chi^2_{4\rm C}$ is retained for further analysis.

\section{Analysis }
\label{sec:ana}

\begin{figure}[htbp]
\begin{center}
\includegraphics[width=8.5cm]{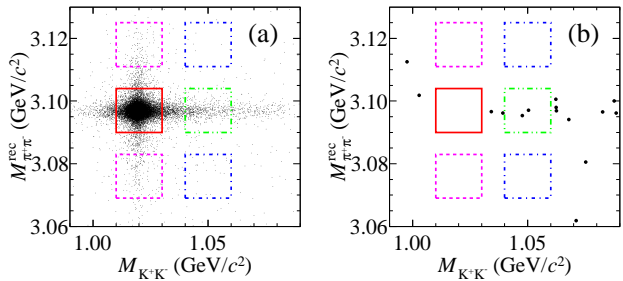}
\caption{Distribution of $ M^{\rm rec}_{\pi^+\pi^-} $ versus $M_{K^+K^-}$ from (a) signal MC sample and (b) $\psi(3686)$ data.
The signal region is defined as the solid box, and the three types of sideband regions (described in the text) are represented by the dashed,
 dashed double dotted, and dashed dotted boxes.
}\label{fig:datasig}
\end{center}
\end{figure}

\begin{figure}[htbp]
\begin{center}
 \includegraphics[width=8.5cm]{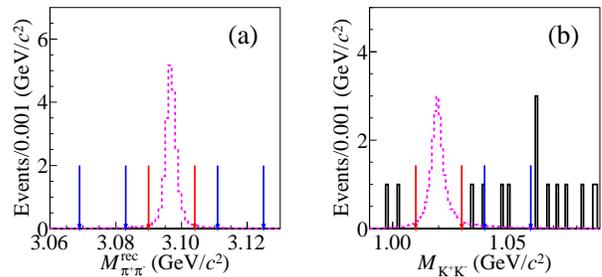}
\caption{Distributions of (a) $ M^{\rm rec}_{\pi^+\pi^-} $ and (b) $M_{K^+K^-}$.
The solid histograms represent the $\psi(3686)$ data and the dashed histograms represent signal MC with arbitrary scale.
The region between the red arrows denotes the signal region, and the one(s) between the blue arrows denotes the
sideband region(s).  }\label{fig:datapro}
\end{center}
\end{figure}

The process $J/\psi \to \phi e^+e^-$ is studied by examining the two-dimensional distribution of the
 $M_{\pi^+\pi^-}^{\rm rec}$ versus the invariant mass of the $K^+K^-$ pair, $M_{K^+K^-}$.
Figure~\ref{fig:datasig}(a) shows the distribution for the signal MC sample,
 where the signal region (shown as a red solid box) is defined as $|M_{K^+K^-}-M_\phi|<0.010 $~GeV/$c^2$ and
$ |M^{\rm rec}_{\pi^+\pi^-}-M_{J/\psi}|<0.007 $~GeV/$c^2$,
where $M_\phi$ and $M_{J/\psi}$ are the nominal masses of $\phi$ and $J/\psi$ mesons
in taken from PDG~\cite{pdg18}, respectively.
The five boxes with equal area around the signal region
are selected as sideband regions, which are categorized into three types.
The first type is used to estimate the background without a $J/\psi$ in the intermediate state;
the second one is for the estimation of the background without a $\phi$ in the intermediate state.
These first two are shown as the pink dashed boxes
and the green dashed double dotted box, respectively.
The third type is for the estimation of the background that includes neither a $J/\psi$ nor a $\phi$ in the intermediate state,
and is shown as blue dashed dotted boxes.
Figure~\ref{fig:datasig}(b) shows the corresponding plot for the $\psi(3686)$ data sample.
No events are observed in the signal region and two events are observed in the $\phi$ sideband.
The non-flat non-$\phi$ background, mainly due to the threshold effect, is estimated by
$\psi(3686)\to \pi^+\pi^- J/\psi $ with $J/\psi \to \phi \pi^+\pi^-$ and
$\phi \to K^+K^-$. The scale factor is determined to be $0.8$ for the background in the $\phi$ signal region to
that in the $\phi$ sideband region. Therefore, the scaled background estimated by the sideband data is $1.6\pm1.1$ events.
The projections of Fig.~\ref{fig:datasig}(b) on $ M^{\rm rec}_{\pi^+\pi^-} $ and $M_{K^+K^-}$
are shown in Figs.~\ref{fig:datapro}(a) and (b), respectively.

The backgrounds from $\psi(3686)$ decays are also studied with an inclusive MC sample of 506 million $\psi(3686)$ events.
No events survive in the signal region, and only one event is found in the second type of sideband region.
This event is from $\psi(3686) \to \pi^+\pi^- J/\psi, J/\psi \to K^+K^- \pi^0, \pi^0 \to \gamma e^+e^-$,
which
will not form a peak in $\phi$ signal region.
In addition, the potential peaking backgrounds from
$\psi(3686) \to \pi^+\pi^-J/\psi$ with $J/\psi \to \phi \eta/\pi^0$ and $\eta/\pi^0 \to \gamma e^+e^-$ are studied
through exclusive MC events generated with a size corresponding to more than
100 times of that of data. The contribution from these channels is negligible.

Possible background sources from continuum processes are estimated with
$44$~pb$^{-1}$ of data collected at a center-of-mass energy
$\sqrt{s} = 3.65$~GeV~\cite{condata}, which is about one fifteenth of the integrated
luminosity of the $\psi(3686)$ data. There are no events satisfying the above selection criteria, therefore we neglect the continuum background.

\section{Systematic Uncertainties}
\label{syserr}

The systematic uncertainties
originate mainly from the
number of $\psi(3686)$ events, the tracking efficiency, the PID efficiency,
the kinematic fit, the selection of the $J/\psi$ and $\phi$ signal regions,
background estimation,
MC statistics, and the branching fractions of intermediate decays. These are
discussed in detail in the following,
and are summarized in Table~\ref{tab:syssum}.

The uncertainty from the total number of $\psi(3686)$ events is estimated
to be $0.7$\%~\cite{npsip}.

The tracking efficiencies for $\pi^{\pm}$ mesons
have been studied with the process $\psi(3686) \to \pi^+\pi^- J/\psi$, $ J/\psi\to \ell^+\ell^-(\ell=e,\mu)$. The difference in
the efficiencies between data and MC simulation is $1.0$\% per pion~\cite{ktrkpid}.
The tracking efficiencies for $K^\pm$ mesons as functions of transverse momentum have been studied
with the process $J/\psi \to K_S^0K^\pm\pi^\mp$, $K_S^0 \to \pi^+\pi^-$. The difference in
the efficiencies between data and MC simulation is $1.0$\% per kaon~\cite{ktrkpid}.
The tracking efficiencies for $e^\pm$ are obtained with a control sample of radiative
Bhabha scattering $e^+e^- \to \gamma e^+e^- $ (including $J/\psi\to \gamma e^+e^- $ ) at
the $J/\psi$ resonance in Ref.~\cite{etrk}. The difference in tracking efficiencies between data and MC
simulation is calculated bin-by-bin over the distribution of transverse momentum versus the polar angle of the lepton
tracks. The uncertainty is determined to be $1.0$\% per electron/positron. The systematic uncertainties arising from
the different charged tracks are summed linearly to be $6.0$\%.

High purity control samples of $e^+e^- \to \gamma e^+e^-$ and $J/\psi \to K^+K^- \pi^0$ have been
selected to study the electron/positron and kaon PID uncertainty. The difference of PID efficiency between data and MC simulation is
calculated in bins of momentum and cos$\theta$. Averaged systematic
uncertainties for electron/positron and kaon identification are obtained by weighting the
difference with the events in each bin of momentum and cos$\theta$ from the signal MC sample, and
determined to be $0.4$\% per electron/positron and $0.8$\% per kaon. Adding these values linearly,
the PID systematic uncertainty is determined to be $2.4$\%.

The systematic uncertainty of the $4$C kinematic fit is studied using a control sample
of $\psi(3686) \to \pi^+\pi^- J/\psi$ with $J/\psi \to \phi \pi^+\pi^-$, $\phi \to K^+K^-$.
The efficiency difference between data and MC simulation with the $\chi^2_{\rm 4C} < 40$ requirement is $3.3$\%,
which is assigned as the systematic uncertainty.

The uncertainty from the signal regions of $M_{K^+K^-}$ and $ M^{\rm rec}_{\pi^+\pi^-} $,
due to their resolution difference between data and MC simulation, is
studied by means of the control sample $\psi(3686) \to \pi^+\pi^- J/\psi$,
$ J/\psi\to \phi \pi^+\pi^-$, $\phi \to K^+K^-$.
The efficiency differences in the $M_{K^+K^-}$ and $ M^{\rm rec}_{\pi^+\pi^-} $
signal regions between data and MC simulation
are $0.2$\% and $1.8$\%, respectively. Adding them in quadrature yields $1.8$\%,
which is taken as the systematic uncertainty.

The uncertainty on the background estimation is studied by an alternative estimation of the
scale factor~($0.74$), estimated using the inclusive MC sample instead of data. The
difference between the resulting upper limits
is taken as the systematic uncertainty, which is $1.5$\%.

The uncertainty of the detection efficiency attributed to the limited size of the MC sample,
$0.4$\%, is taken as the systematic uncertainty from MC statistics.

To estimate the uncertainty from the model used to simulate the $J/\psi \to \phi e^+e^-$ decay,
we generated a MC sample based on the phase-space assumption.
The difference between the efficiencies determined from the MC sample described in Sec.~\ref{sec:bes3}
and the phase space MC sample is $10.7$\%. We take
half of the difference ($5.4$\%) as the systematic uncertainty from MC modeling.

In the determination of the upper limit on the branching fraction of the
process of interest, we have accounted for the branching fractions of
$\psi(3686) \to \pi^+\pi^- J/\psi$ and $\phi \to K^+K^-$ by taking the values given by the
PDG~\cite{pdg18}. The uncertainties of these cited values
are taken as a source of systematic uncertainty, which are $0.9$\% and $1.0$\%, respectively.

The total systematic uncertainty, $\Delta_{\rm sys}$, is calculated by adding the uncertainties from all sources in quadrature.

\begin{table}[htbp]
\begin{center}
\caption{ Summary of relative systematic uncertainties (in \%). }\label{tab:syssum}
\begin{small}
\begin{tabular}{lcc}\hline
Sources            & Uncertainty                \\ \hline
$N^{\rm tot}_{\psi(3686)}$   & 0.7            \\
Tracking           & 6.0                 \\
PID                & 2.3                      \\
Kinematic fit      & 3.3                       \\
Signal region      & 1.8                     \\
Background estimation      & 1.5             \\
MC statistics      & 0.4                      \\
MC modeling        & 5.4                      \\
$\mathcal{B}(\psi(3686) \to \pi^+\pi^- J/\psi)$   & 0.9   \\
$\mathcal{B}(\phi\to K^+K^-)$  & 1.0   \\ \hline
Total              & 9.5                       \\
\hline
\end{tabular}
\end{small}
\end{center}
\end{table}

\section{ \bf Result}

Since no candidate events are observed in the signal region and $1.6\pm1.1$ background events are estimated,
the upper limit on the number of $\psi(3686) \to \pi^+\pi^- J/\psi$ events with
$J/\psi \to \phi e^+e^-$ is set to be $1.31$ at the $90$\% confidence level (CL) using the
Feldman-Cousins~\cite{feldman} method with the assumption of a Poisson process.

After taking into account the systematic uncertainty~\cite{cousins}, the upper
limit of the branching fraction of $J/\psi \to \phi e^+e^-$ is calculated with
\begin{equation} \label{eq:bfup}
\mathcal{B} (J/\psi \to \phi e^+e^-) < \frac{N^{\rm up}\times(1+N^{\rm up}\times
\Delta_{\rm sys}^2/2)}{N^{\rm tot}_{\psi(3686)} \times \epsilon
\times \prod \mathcal{B}_i },
\end{equation}
where $N^{\rm tot}_{\psi(3686)}$ is the
number of $\psi(3686)$ events, $\prod \mathcal{B}_i$ represents the branching
fraction product $\mathcal{B}(\psi(3686) \to \pi^+\pi^- J/\psi) \times \mathcal{B}(\phi\to K^+K^-)$ and $\epsilon$
is the detection efficiency, which is $(14.31\pm0.05)$\% determined by MC simulations as described in
Sec.~\ref{sec:bes3}.
Table~\ref{tab:results} summarizes the various values that were used as input to Eq.~(\ref{eq:bfup}).
We find an upper limit on the branching fraction of the $J/\psi \to \phi e^+e^-$ process at the 90\% CL of
\begin{equation}
\mathcal{B} (J/\psi \to \phi e^+e^-) <1.2\times 10^{-7}. \nonumber
\end{equation}

\begin{table}[htbp]
\begin{center}
\caption{ Input values used to obtain the upper
limit on the branching fraction of $J/\psi \to \phi e^+e^-$. $N^{\rm obs}$, $N^{\rm bkg}$
and $N^{\rm up}$ represent the number of
observed events, background events, and the upper limit on the number of observed events,
respectively. }\label{tab:results}
\begin{small}
\begin{tabular}{lr}\hline
Item                           & Value     \\ \hline
$N^{\rm tot}_{\psi(3686)} $    & $(448.1\pm 2.9) \times 10^{6}$    \\
$N^{\rm obs}$                  & $0$      \\
$N^{\rm bkg}$                  & $1.6\pm1.1$    \\
$N^{\rm up}$(90\% CL)        & $1.31$    \\
$\epsilon$                     & $(15.13\pm0.05)$\%    \\  
$\mathcal{B}(\psi(3686) \to \pi^+\pi^- J/\psi)$ & $(34.49\pm0.30)$\% \\
$\mathcal{B}(\phi \to K^+K^-)$ & $(48.9\pm0.5)$\% \\
$\Delta_{\rm sys}$             & 9.5\%             \\ \hline
$ \mathcal{B}(J/\psi \to \phi e^+e^-)$       & $<1.2\times 10^{-7}$   \\
\hline
\end{tabular}
\end{small}
\end{center}
\end{table}

\section{ \bf Summary}

Using the $448.1\times10^{6}$ $\psi(3686)$ events collected with the BESIII detector,
we report a search for the rare decay $J/\psi \to \phi e^+e^-$
via $\psi(3686) \to \pi^+\pi^- J/\psi$. No
signal events are observed and the upper limit of the branching fraction
for this decay is calculated to be $\mathcal{B} (J/\psi \to \phi e^+e^-) <1.2\times 10^{-7} $
by the Feldman-Cousins
method at the $90$\% CL,
which is one order of magnitude higher than the prediction in Ref.~\cite{guoxd}. Our result shows that,
even if there are new particles involved in the Fig.~\ref{fig:feynman}(b) process, their
contributions will not be too large, and the masses of the possible existing
new particles or their coupling information to vector mesons ($J/\psi$ and $\phi$)
can also be constrained.

\section{Acknowledgements}

The BESIII collaboration thanks the staff of BEPCII and the IHEP computing center for
their strong support. This work is supported in part by National Key Basic Research
Program of China under Contract No. 2015CB856700; National Natural Science Foundation
of China (NSFC) under Contracts Nos. 11575077, 11475090, 11235011, 11335008, 11425524,
11625523, 11635010, 11735014, 11835012; the Outstanding
Youth project of Natural Science Foundation of Hunan Province; the Chinese Academy of
Sciences (CAS) Large-Scale Scientific Facility Program; Joint Large-Scale Scientific
Facility Funds of the NSFC and CAS under Contracts Nos. U1532257, U1532258, U1732263,
U1832207; CAS Key Research Program of Frontier Sciences under Contracts Nos. QYZDJ-SSW-SLH003,
QYZDJ-SSW-SLH040; 100 Talents Program of CAS; INPAC and Shanghai Key Laboratory for
Particle Physics and Cosmology; German Research Foundation DFG under Contract No.
Collaborative Research Center CRC 1044; Istituto Nazionale di Fisica Nucleare, Italy;
Koninklijke Nederlandse Akademie van Wetenschappen (KNAW) under Contract No. 530-4CDP03;
Ministry of Development of Turkey under Contract No. DPT2006K-120470; National Science
and Technology fund; The Knut and Alice Wallenberg Foundation (Sweden) under Contract
No. 2016.0157; The Swedish Research Council; U. S. Department of Energy under Contracts
Nos. DE-FG02-05ER41374, DE-SC-0010118, DE-SC-0012069; University of Groningen (RuG) and
the Helmholtzzentrum f\"{u}r Schwerionenforschung GmbH (GSI), Darmstadt.

\end{document}

%% file: authors_aug2017.tex
\author{
\begin{small}
\begin{center}
M.~Ablikim$^{1}$, M.~N.~Achasov$^{10,d}$, S. ~Ahmed$^{15}$, M.~Albrecht$^{4}$, M.~Alekseev$^{55A,55C}$, A.~Amoroso$^{55A,55C}$, F.~F.~An$^{1}$, Q.~An$^{52,42}$, Y.~Bai$^{41}$, O.~Bakina$^{27}$
% ...
\\
\vspace{0.2cm}
(BESIII Collaboration)\\
\vspace{0.2cm} {\it
$^{1}$ Institute of High Energy Physics, Beijing 100049, People's Republic of China\\
% ...
$^{4}$ Bochum Ruhr-University, D-44780 Bochum, Germany\\
% ...
$^{10}$ G.I. Budker Institute of Nuclear Physics SB RAS (BINP), Novosibirsk 630090, Russia\\
% ...
$^{15}$ Helmholtz Institute Mainz, Johann-Joachim-Becher-Weg 45, D-55099 Mainz, Germany\\
% ...
$^{27}$ Joint Institute for Nuclear Research, 141980 Dubna, Moscow region, Russia\\
% ...
$^{41}$ Southeast University, Nanjing 211100, People's Republic of China\\
$^{42}$ State Key Laboratory of Particle Detection and Electronics, Beijing 100049, Hefei 230026, People's Republic of China\\
% ...
$^{52}$ University of Science and Technology of China, Hefei 230026, People's Republic of China\\
% ...
$^{55}$ (A)University of Turin, I-10125, Turin, Italy; (B)University of Eastern Piedmont, I-15121, Alessandria, Italy; (C)INFN, I-10125, Turin, Italy\\
% ...
$^{d}$ Also at the Novosibirsk State University, Novosibirsk, 630090, Russia\\
% ...
}\end{center}
\vspace{0.4cm}
\end{small}
}

\affiliation{}

%% file: draft-phiee.bbl
\begin{thebibliography}{99}
\bibitem{guoxd} X.~D. Guo, S.~R. Xue, H.~W. Ke, X.~Q. Li, Q. Zhao, Chin.\ Phys.\ C {\bf 40}, 073104 (2016).
\bibitem{bes3} M.~Ablikim {\it et al.} [BESIII Collaboration], Nucl.\ Instrum.\ Meth.\ A {\bf 614}, 345 (2010).
\bibitem{bes2} J.~Z.~Bai {\it et al.} [BES Collaboration], Nucl.\ Instrum.\ Meth.\ A {\bf 344}, 319 (1994); {\bf 458}, 627 (2001).
\bibitem{geant4} S.~Agostinelli {\it et al.} [GEANT4 Collaboration], Nucl.\ Instrum.\ Meth.\ A {\bf 506}, 250 (2003).
\bibitem{boost} Z.~Y.~Deng {\it et al.,} High Energy Physics \& Nuclear Physics {\bf 30}, 371 (2006).
\bibitem{kkmc} S.~Jadach, B.~F.~L.~Ward, and Z.~Was, Comput.\ Phys.\ Commun.\ {\bf 130}, 260 (2000); Phys.\ Rev.\ D {\bf 63}, 113009 (2001).
\bibitem{evtgen} R.~G.~Ping, Chin.\ Phys.\ C {\bf 32}, 599 (2008); D.~J.~Lange, Nucl.\ Instrum.\ Meth.\ A {\bf 462}, 152 (2001).
\bibitem{pdg18} M.~Tanabashi {\it et al.} [Particle Data Group], Phys.\ Rev.\ D {\bf 98}, 030001 (2018).
\bibitem{lundcharm} J.~C.~Chen, G.~S.~Huang, X.~R.~Qi, D.~H.~Zhang, Y.~S.~Zhu, Phys. \ Rev.\ D \ {\bf 62}, 034003 (2000);
R.~L.~Yang, R.~G.~Rong and D.~Chen, Chin.\ Phys.\ \ Lett. {\bf 31}, 061301 (2014).
\bibitem{bes2jpsipipi} M.~Ablikim {\it et al.} [BES Collaboration], Phys.\ Rev.\ D {\bf 62}, 032002 (2000).
\bibitem{lifengyun} M.~Ablikim {\it et al.} [BESIII Collaboration], arxiv:1809.00635.
\bibitem{condata} M.~Ablikim, {\it et al.} [BESIII Collaboration], Chin.\ Phys.\ C \ {\bf 37}, 123001 (2013).
\bibitem{npsip} M.~Ablikim {\it et al.} [BESIII Collaboration], Chin.\ Phys.\ C \ {\bf 42}, 023001 (2018).
\bibitem{ktrkpid} M.~Ablikim {\it et al.} [BESIII Collaboration], Phys.\ Rev.\ D {\bf 83}, 112005 (2011).
\bibitem{etrk} M.~Ablikim {\it et al.} [BESIII Collaboration], Phys.\ Rev.\ D {\bf 89}, 092008 (2014).
\bibitem{feldman} G.~J.~Feldman and R.~D.~Cousins, Phys.\ Rev.\ D {\bf 57}, 3873 (1998).
\bibitem{cousins} R.~D.~Cousins and V.~L.~Highland, Nucl.\ Instrum.\ Meth.\ A {\bf 320}, 331 (1992).
\end{thebibliography}
